\documentclass[a4paper]{scrartcl}

\usepackage[T1]{fontenc}
\usepackage[utf8]{inputenc}
\usepackage{acro}
\usepackage{tikz}
\usepackage{units}
\usepackage{wrapfig}
\usepackage{hyperref}
\usepackage[binary-units=true]{siunitx}
\usepackage[ruled,vlined,linesnumbered,noresetcount]{algorithm2e}
\usepackage{graphicx}
\usepackage{acro}
\usepackage{amsthm}
\usepackage{amsmath}
\usepackage{amssymb}
\usepackage{cite}
\usepackage{caption}
\usepackage{subcaption}
\usepackage{cleveref}
\usepackage{authblk}
\usepackage{xpatch}
\usepackage[bottom=3cm]{geometry}

\newtheorem{theorem}{Theorem}[section]
\newtheorem{definition}[theorem]{Definition}
\newtheorem{lemma}[theorem]{Lemma}

\crefname{algocf}{alg.}{algs.}
\Crefname{algocf}{Algorithm}{Algorithms}

\SetAlFnt{\footnotesize}

\graphicspath{{figs/}{plots/}}

\title{An Efficient, Practical Algorithm and Implementation for Computing Multiplicatively Weighted Voronoi Diagrams}

\author[1]{Martin Held}
\author[1]{Stefan de Lorenzo}

\affil[1]{Universit\"at Salzburg, FB Computerwissenschaften, Austria}
\affil[ ]{\texttt{\{held,slorenzo\}@cs.sbg.ac.at}}

\date{}

\setkomafont{title}{\normalfont\bfseries}
\makeatletter
\patchcmd{\@maketitle}{\huge}{\LARGE}{}{}
\makeatother

\newcommand{\BigO}{\mathcal{O}}
\newcommand{\R}{\mathbb{R}}

\newcommand{\Dist}[3]{d_{#1}(#2,#3)}
\newcommand{\Sort}{\mathcal{L}}

\newcommand{\VorReg}[2]{\mathcal{VR}_w(#1,#2)}
\newcommand{\VorDiag}[1]{\mathcal{VD}_w(#1)}
\newcommand{\Wavefr}[1]{\mathcal{WF}(#1)}
\newcommand{\ArcArr}[1]{\mathcal{A}(#1)}
\newcommand{\Prio}{\mathcal{Q}}

\newcommand{\Off}[2]{c_{#1}(#2)}

\newcommand{\Vertex}[1]{v_{#1}}

\newcommand{\ActArc}[1]{a_{#1}}
\newcommand{\Bisec}[2]{b_{#1,#2}}
\newcommand{\ActOff}[1]{o_{#1}}

\newcommand{\OvArr}{\mathcal{O}\mathcal{A}}
\newcommand{\Dcel}{\mathcal{D}}
\newcommand{\SalzburgDB}{Salzburg database of polygonal data }

\DeclareAcronym{mwvd}{
  short = MWVD,
  long = multiplicatively weighted Voronoi diagram,
  short-plural = s,
  long-plural = s
}

\DeclareAcronym{awvd}{
  short = AWVD,
  long = additively weighted Voronoi diagram,
  short-plural = s,
  long-plural = s
}

\DeclareAcronym{aa}{
  short = AA,
  long = arc arrangement,
  short-plural = s,
  long-plural = s
}

\DeclareAcronym{cgal}{
  short = CGAL,
  long = the Computational Geometry Algorithms Library
}

\begin{document}

\maketitle

\begin{abstract}
  We present a simple wavefront-like approach for computing
  multiplicatively weighted Voronoi diagrams of points and
  straight-line segments in the Euclidean plane. If the input sites may be
  assumed to be randomly weighted points then the use of a so-called overlay
  arrangement [Har-Peled\&Raichel, Discrete Comput.\ Geom.~53:547--568, 2015]
  allows to achieve an expected runtime complexity of $\BigO(n\log^4 n)$,
  while still maintaining the simplicity of our approach.  We implemented the
  full algorithm for weighted points as input sites, based on CGAL. The
  results of an experimental evaluation of our implementation suggest $\BigO(n
  \log^2 n)$ as a practical bound on the runtime.
  Our algorithm can be extended to handle also additive weights in addition to
  multiplicative weights, and it yields a truly simple $\BigO(n \log
  n)$ solution for solving the one-dimensional version of this problem.
\end{abstract}

\section{Introduction}

The \ac{mwvd} was introduced by Boots \cite{boots1980weighting}.
Aurenhammer and Edelsbrunner~\cite{aurenhammer1984optimal} present a
worst-case optimal incremental algorithm for constructing the \ac{mwvd} of a
set of $n$ points in $\BigO(n^2)$ time and space. They define spheres on the
bisector circles (that are assumed to be situated in the $xy$-plane) and
convert them into half-planes in $\R^3$ using a spherical inversion.
Afterwards, these half-planes are intersected. Thus, every Voronoi region is
associated with a polyhedron. Finally, the intersection of every such
polyhedron with a sphere that corresponds to the $xy$-plane is inverted back
to $\R^2$. We are not aware of an implementation of their algorithm, though.
(And it seems difficult to implement.) In any case, the linear-time repeated
searches for weighted nearest points indicate that its complexity is
$\Theta(n^2)$ even if the combinatorial complexity of the resulting Voronoi
diagram is $o(n^2)$.  Later Aurenhammer uses divide\&conquer to obtain an
$\BigO(n \log n)$ time and $\BigO(n)$ space algorithm for the one-dimensional
weighted Voronoi diagram~\cite{aurenhammer1986one}.

Har-Peled and Raichel \cite{har2015complexity} show that a bound of $\BigO(n
\log^2 n)$ holds on the expected combinatorial complexity of a \ac{mwvd} if
all weights are chosen randomly. They sketch how to compute \acp{mwvd} in
expected time $\BigO(n \log^3 n)$. Their approach is also difficult to implement
because it uses the algorithm by Aurenhammer and
Edelsbrunner~\cite{aurenhammer1984optimal} as a subroutine.

Vyatkina and Barequet \cite{vyatkina2011multiplicatively} present a
wavefront-based strategy to compute the \ac{mwvd} of a set of $n$ lines in the
plane in $\BigO(n^2\log n)$ time. The Voronoi nodes are computed based on edge
and break-through events.  An edge event takes place when an wavefront edge
disappears. A break-through event happens whenever a new wavefront edge
appears.

Since the pioneering work of Hoff et al.~\cite{hoff1999fast} it has been well
known that discretized versions of Voronoi diagrams can be computed using the
GPU framebuffer. More recently, Bonfiglioli et al.~\cite{Bon*14} presented
a refinement of this rendering-based approach. It is obvious that their
approach could also be extended to computing approximate \acp{mwvd}.  However,
the output of such an algorithm is just a set of discrete pixels instead of a
continuous skeletal structure. Its precision is limited by the resolution of
the framebuffer and by the numerical precision of the depth buffer.

\section{Our Contribution}

Our basic algorithm allows us to compute \acp{mwvd} in worst-case
$\BigO(n^2\log n)$ time and $\BigO(n^2)$ space. A refined version makes use of
the result by Har-Peled and Raichel \cite{har2015complexity}: We use their
overlay arrangement to keep the expected runtime complexity bounded by
$\BigO(n \log^4 n)$ if the point sites are weighted randomly.  Hence, for the
price of a multiplicative factor of $\log n$ we get an algorithm that is easier to implement. Our experiments suggest that this bound is too pessimistic in
practice and that one can expect the actual runtime to be bounded by $\BigO(n
\log^2 n)$. However, our experiments also show that one may get a quadratic
runtime if the weights are not chosen randomly. Our algorithm does not
require the input sites to have different multiplicative weights, and it can
be extended to additive weights and to (disjoint) straight-line segments as
input sites. Furthermore, it yields a truly simple $\BigO(n \log n)$ solution
for computing \acp{mwvd} in one dimension, where all input points lie on a
line.

Our implementation is based on exact arithmetic and \ac{cgal}~\cite{CGAL}.
It is publicly available on GitHub under \url{https://github.com/cgalab/wevo}.  To the best of our knowledge,
this is the first full implementation of an algorithm for computing \acp{mwvd}
that achieves a decent expected runtime complexity.

\section{Preliminaries}

Let $S:=\{s_1,s_2,\ldots,s_n\}$ denote a 
set of $n$ distinct weighted points in $\R^2$ that are indexed such that
$w(s_i)\leq w(s_j)$ for $1\le i<j\le n$, where $w(s_i)\in\R^+$ is the
weight associated with $s_i$. It is common to regard the weighted
distance $\Dist{w}{p}{s_i}$ from an arbitrary point $p$ in $\R^2$ to $s_i$ as
the standard Euclidean distance $\Dist{}{p}{s_i}$ from $p$ to $s_i$ divided by
the weight of $s_i$, i.e., $d_w(p,s_i) := \frac{\Dist{}{p}{s_i}}{w(s_i)}$.
The \emph{(weighted) Voronoi region} $\VorReg{s_i}{S}$ of $s_i$ relative to
$S$ is the set of all points of the plane such that no site $s_j$ in
$S\setminus\{s_i\}$ is closer to $p$ than $s_i$, that is,
$\VorReg{s_i}{S}:=\left\{p \in \R^2 : d_w(p,s_i) \leq d_w(p,s_j)\text{~for
    all~} j\in\{1,2,\ldots,n\}\right\}$.  Then the \acf{mwvd}, $\VorDiag{S}$,
of $S$ is defined as $\VorDiag{S} := \bigcup_{s_i\in S}
\partial\,\VorReg{s_i}{S}$.

\begin{figure}[htbp]
\centering
\includegraphics[page=3]{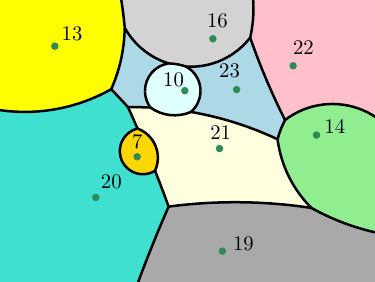}
\hspace{.3cm}
\includegraphics[page=4]{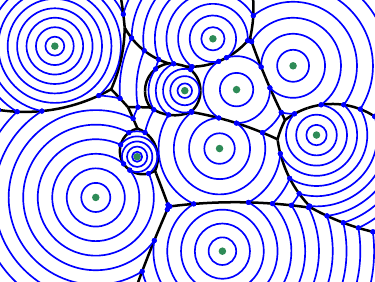}
\caption{Left: The numbers next to the points indicate their weights and the
  corresponding \ac{mwvd} is shown. Right: Wavefronts (in blue) for equally-spaced points in time.}
\label{fig:mwvd}
\end{figure}

A connected component of a Voronoi region is called a \emph{face}. For two
distinct sites $s_i$ and $s_j$ of $S$, the \emph{bisector} $\Bisec{i}{j}$ of
$s_i$ and $s_j$ models the set of points of the plane that are at the same
weighted distance from $s_i$ and $s_j$. Hence, a non-empty intersection of two
Voronoi regions is a subset of the bisector of the two defining sites.
Following common terminology, a connected component of such a set is called a
\emph{(Voronoi) edge} of $\VorDiag{S}$. An end-point of an edge is called a
\emph{(Voronoi) node}. It is known that the bisector between two unequally
weighted sites forms a circle\footnote{Apollonius of Perga defined a circle as
  a set of points that have a specific distance ratio to two foci.}.
An example of a \ac{mwvd} is shown in \Cref{fig:mwvd}.

The wavefront $\Wavefr{S,t}$ emanated by $S$ at time $t\ge 0$ is the set of
all points $p$ of the plane whose minimal weighted distance from $S$ equals
$t$. More formally,
\[
\Wavefr{S,t} := \left\{p\in\R^2 : \min_{s_i\in S} d_w(p,s_i)=t\right\}.
\]
The wavefront consists of circular arcs which we call
\emph{wavefront arcs}. A common end-point of two consecutive wavefront arcs is
called \emph{wavefront vertex}; see the blue dots in \Cref{fig:mwvd}.

\section{Offset Circles}
 
For the sake of descriptional simplicity, we 
start with assuming that no point in the plane has the same weighted distance
to more than three sites of $S$.
For $t\ge 0$, the \emph{offset circle} $\Off{i}{t}$ of the $i$-th site $s_i$
is given by a circle centered at $s_i$ with radius $t\cdot w(s_i)$. We find it
convenient to regard $\Off{i}{t}$ as a function of either time or distance
since at time $t$ every point on $\Off{i}{t}$ is at Euclidean distance $t
\cdot w(s_i)$ from $s_i$, i.e., at weighted distance $t$. 
We specify a point of $\Off{i}{t}$ relative to $s_i$ by its polar angle
$\alpha$ and its (weighted) polar radius $t$ and denote it by $p_i(\alpha,t)$.

\begin{figure}[htbp]
\centering
\begin{subfigure}{.24\textwidth}
\includegraphics[page=3]{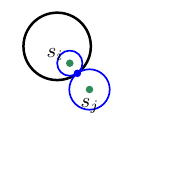}
\end{subfigure}
\begin{subfigure}{.24\textwidth}
\includegraphics[page=4]{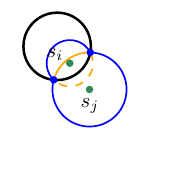}
\end{subfigure}
\begin{subfigure}{.24\textwidth}
\includegraphics[page=6]{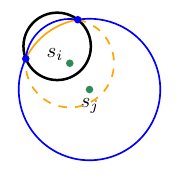}
\end{subfigure}
\begin{subfigure}{.24\textwidth}
\includegraphics[page=7]{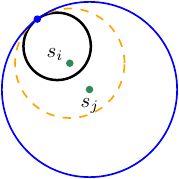}
\end{subfigure}
\caption{Two married vertices (highlighted by the blue dots) trace out the
  bisector $b_{ij}$ (in black).}
\label{fig:traj}
\end{figure}

For $1\le i < j\le n$, consider two sites $s_i,s_j\in S$ and assume that
$w(s_i)\neq w(s_j)$. Then there exists a unique closed time interval
$[t_{ij}^{min},t_{ij}^{max}]$ during which the respective offset circles of
$s_i,s_j$ intersect. We say that the two offset circles \emph{collide} at
their mutual \emph{collision time} $t_{ij}^{min}$, and $s_j$ starts to
\emph{dominate} $s_i$ at the domination time $t_{ij}^{max}$. 
For all other times $t$ within this interval the two
offset circles $\Off{i}{t}$ and $\Off{j}{t}$ intersect in two disjoint points
$\Vertex{i,j}^l(t)$ and $\Vertex{i,j}^r(t)$. These \emph{(moving) vertices}
trace out the bisector between $s_i$ and $s_j$; see \Cref{fig:traj}. Since
$\Vertex{i,j}^l(t)$ and $\Vertex{i,j}^r(t)$ are defined by the same pair of
offset circles we refer to $\Vertex{i,j}^l(t)$ as the \emph{vertex married to}
$\Vertex{i,j}^r(t)$, and vice versa. Every other pair of moving vertices
defined by two different pairs of intersecting offset circles is called
\emph{unmarried}.  To simplify the notation, we will drop the parameter $t$ if
we do not need to refer to a specific time. Similarly, we 
drop the superscripts $l$ and $r$ if no distinction between married and
unmarried vertices is necessary.

\section{A Simple Event-Based Construction Scheme}

\label{sec:event_based}
In this section we describe a simulation of a propagation of the wavefront
$\Wavefr{S,t}$ 
to compute $\VorDiag{S}$. Since the wavefront is given by a subset of the arcs
of the arrangement of all offset circles, one could attempt to study the
evolution of all arcs of that arrangement over time.  However, it is
sufficient to restrict our attention to a subset of arcs of that arrangement.
We note that our wavefront can be seen as a kinetic data structure
\cite{Guib01}. 

Clearly, the arc along $\Off{i}{t}$ which is inside $\Off{j}{t}$ will not
belong to $\Wavefr{\{s_i,s_j\},t^*}$ for any $t^*>t$. We will make use of
this observation to define \emph{inactive} and \emph{active arcs} that are
situated along the offset circles. 

\begin{definition}[Active point] \label{def:active_point}
  A point $p$ on the offset circle $\Off{i}{t}$ is called \emph{inactive} at
  time $t$ 
  (relative to $S$) if there exists $j>i$, with $1\le i < j\le n$, such that
  $p$ lies strictly inside of $\Off{j}{t}$. Otherwise, $p$ is \emph{active}
  (relative to $S$) at time $t$. A vertex $\Vertex{i,j}(t)$ is an
  \emph{active vertex} if it is an active point on both $\Off{i}{t}$ and
  $\Off{j}{t}$ at time $t$; otherwise, it is an \emph{inactive vertex}.
\end{definition}

\begin{lemma}
  If $p_i(\alpha,t)$ is inactive at time $t$ then $p_i(\alpha,t')$ will
  be inactive for all times $t'\ge t$.\label{lem:act_point}
\end{lemma}

An inactive point $p_i(\alpha,t)$ cannot be part of the wavefront
$\Wavefr{S,t}$. \Cref{lem:act_point} ensures that none of its future
incarnations $p_i(\alpha,t')$ can become part of the wavefront
$\Wavefr{S,t'}$.

\begin{definition}[Active arc]
  For $1\le i \le n$ and $t\ge 0$, an \emph{active arc} of the offset circle
  $\Off{i}{t}$ at time $t$ is a maximal connected set of points on
  $\Off{i}{t}$ that are active at time $t$. The closure of a maximal connected
  set of inactive points of $\Off{i}{t}$ forms an \emph{inactive arc} of
  $\Off{i}{t}$ at time $t$.\label{def:act_arc}
\end{definition}

Every end-point of an active arc of $\Off{i}{t}$ is given by the intersection
of $\Off{i}{t}$ with some other offset circle $\Off{j}{t}$, i.e., by a moving
vertex $\Vertex{i,j}(t)$. This vertex is active, too. 

\begin{definition}[Arc arrangement]
  The \emph{\acf{aa}} of $S$ at time $t$, $\ArcArr{S,t}$, is the arrangement
  induced by all active arcs of all offset circles of $S$ at time $t$.
\label{def:act_arr}
\end{definition}

As time $t$ increases, the offset circles expand. This causes the vertices of
$\ArcArr{S,t}$ to move, but it will also result in topological changes of
the arc arrangement.

\begin{definition}[Collision event]
  Let $p_i(\alpha,t_{ij}^{min})=p_j(\alpha + \pi,t_{ij}^{min})$ be the point of
  intersection of the offset circles of $s_i$ and $s_j$ at the collision time
  $t_{ij}^{min}$, for some fixed angle $\alpha$. A \emph{collision event}
  occurs between these two offset circles at time $t_{ij}^{min}$ if the points
  $p_i(\alpha,t)$ and $p_j(\alpha + \pi,t)$ have been active for all times $0 \le
  t\le t_{ij}^{min}$.
\end{definition}

At the time of a collision a new pair of married vertices $\Vertex{i,j}^l(t)$
and $\Vertex{i,j}^r(t)$ is created. Of course, we have
$\Vertex{i,j}^l(t_{ij}^{min}) = \Vertex{i,j}^r(t_{ij}^{min}) =
p_i(\alpha,t_{ij}^{min})$.

\begin{definition}[Domination event]
  Let $p_i(\alpha,t_{ij}^{max})=p_j(\alpha,t_{ij}^{max})$ be the point of
  intersection of the offset circles of $s_i$ and $s_j$ at the domination time
  $t_{ij}^{max}$, for some fixed angle $\alpha$. A \emph{domination event}
  occurs between these two offset circles at time $t_{ij}^{max}$ if the points
  $p_i(\alpha,t)$ and $p_j(\alpha,t)$ have been active for all times $0 \le
  t\le t_{ij}^{max}$.
\end{definition}

At the time of a domination event the married vertices
$\Vertex{i,j}^l(t_{ij}^{max})$ and $\Vertex{i,j}^r(t_{ij}^{max})$ coincide and
are removed.

\begin{definition}[Arc event]
  An \emph{arc event} $e$ occurs at time $t_e$ when an active arc
  $\ActArc{i}$ shrinks to zero length because two unmarried vertices
  $\Vertex{i,j}(t_e)$ and $\Vertex{i,k}(t_e)$ meet in a point $p_e$ on
  $\Off{i}{t_e}$.
\end{definition}

\begin{figure}[htb]
  \centering
  \begin{subfigure}{.45\textwidth}
    \includegraphics[page=21]{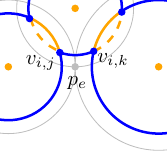}
    \hspace{.1cm}
    \includegraphics[page=22]{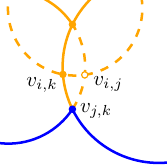}
    \caption{}
    \label{fig:collision_arc_ev}
  \end{subfigure}
  \begin{subfigure}{.45\textwidth}
    \centering
    \includegraphics[page=23]{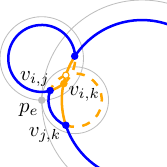}
    \hspace{.1cm}
    \includegraphics[page=24]{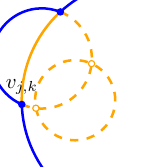}
    \caption{}
    \label{fig:domination_arc_ev}
  \end{subfigure}
  \caption{(a) The configuration shortly before (left) and
    after (right) a collision event as well as an arc
    event occur simultaneously at the same point $p_e$. In the left figure the
    offset arcs at the time of the event are shown in gray. Arcs and vertices
    that are on $\Wavefr{\{s_i,s_j,s_k\},t}$ are highlighted in blue. Other
    active arcs and vertices are depicted by solid orange lines and filled
    disks, while inactive arcs and vertices are depicted by dashed orange
    lines and circles. (b) The configuration shortly before and after a domination event and
    an arc event occur simultaneously at the same point $p_e$.}
\end{figure}

\Cref{lem:act_point} implies that $p_i(\alpha,t)$ has been active for all
times $t\le t_e$ if $p_i(\alpha,t_e)=p_e$.  At the time of an arc event two
unmarried vertices trade their places along an offset circle. 
Now suppose that the two unmarried vertices $\Vertex{i,j}(t_e)$ and
$\Vertex{i,k}(t_e)$ meet in a point $p_e$ along $\Off{i}{t_e}$ at the time
$t_e$ of an arc event, thereby causing an active arc of $\Off{i}{t_e}$ to
shrink to zero length.  Hence, the offset circles of $s_i,s_j$ and $s_k$
intersect at the point $p_e$ at time $t_e$. If $\Off{j}{t}$ and $\Off{k}{t}$
did not intersect for $t<t_e$ then we also get a collision event between
$\Off{j}{t}$ and $\Off{k}{t}$ at time $t_e$, see \Cref{fig:collision_arc_ev}.
(This configuration can occur for any relative order of the weights
$w(s_i),w(s_j),w(s_k)$.) Otherwise, one or both of the married vertices
$\Vertex{j,k}^l(t_e)$ and $\Vertex{j,k}^r(t_e)$ must also coincide with $p_e$.
If both coincide with $p_e$ then we also get a domination event between
$\Off{j}{t}$ and $\Off{k}{t}$ at time $t_e$ and we have $w(s_j)<w(s_k)$, see
\Cref{fig:domination_arc_ev}.  The scenarios remaining for the case that only
one of $\Vertex{j,k}^l(t_e)$ and $\Vertex{j,k}^r(t_e)$ coincides with $p_e$
are detailed in the following lemma.

\begin{figure}[!htb]
  \centering
  \begin{subfigure}{\textwidth}
    \centering
    \includegraphics[page=1]{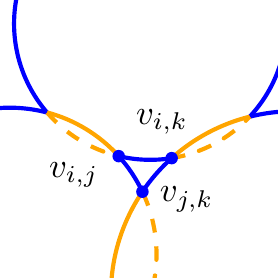}
    \hspace{.1cm}
    \includegraphics[page=2]{mwvd-events}
    \hspace{1cm}
    \includegraphics[page=11]{mwvd-events}
    \hspace{.1cm}
    \includegraphics[page=12]{mwvd-events}
    \caption{The two possible configurations shortly before (shown in the left
      figures) and after (shown in the right figures) one active arc
      disappears on $\Off{i}{t}$ if no collision or domination event occurs at
      the same point. We get the collapse of all three arcs of an active-arc
      triangle.}
    \label{fig:arc_ev_1}
  \end{subfigure}\\

  \begin{subfigure}{\textwidth}
    \centering
    \includegraphics[page=5]{mwvd-events}
    \hspace{.1cm}
    \includegraphics[page=6]{mwvd-events}
    \hspace{1cm}
    \includegraphics[page=9]{mwvd-events}
    \hspace{.1cm}
    \includegraphics[page=10]{mwvd-events}
    \caption{The two possible configurations shortly before (left) and after
      (right) one active arc disappears on $\Off{j}{t}$ and another active arc
      appears on $\Off{k}{t}$.}
    \label{fig:arc_ev_2}
  \end{subfigure}\\

  \begin{subfigure}{\textwidth}
    \centering
    \includegraphics[page=3]{mwvd-events}
    \hspace{.1cm}
    \includegraphics[page=4]{mwvd-events}
    \hspace{1cm}
    \includegraphics[page=7]{mwvd-events}
    \hspace{.1cm}
    \includegraphics[page=8]{mwvd-events}
    \caption{The two possible configurations shortly before (left) and after
      (right) one active arc disappears on $\Off{k}{t}$ and another active arc
      appears on $\Off{j}{t}$.}
    \label{fig:arc_ev_3}
  \end{subfigure}
  \caption{The six different configurations that can occur for arc events for
    $1\le i < j < k \le n$.}
  \label{fig:arc_ev}
\end{figure}

\begin{lemma}
  Let $i<j<k$ and consider an arc event such that exactly the three vertices
  $\Vertex{i,j}(t_e)$, $\Vertex{i,k}(t_e)$, and $\Vertex{j,k}(t_e)$ coincide
  at time $t_e$.  Then either
\begin{itemize}
  \item all three vertices were active before the event, see
    \cref{fig:arc_ev_1}, or 
  \item $\Vertex{i,j}$ and $\Vertex{j,k}$ were active and $\Vertex{i,k}$ was 
    inactive before the event, see \cref{fig:arc_ev_2}, or
  \item $\Vertex{i,k}$ and $\Vertex{j,k}$ were active and $\Vertex{i,j}$ was
    inactive before the event, see \cref{fig:arc_ev_3}.
\end{itemize}
\label{lem:movisects}
\end{lemma}

\begin{figure}[!htb]
\centering
\includegraphics[page=18]{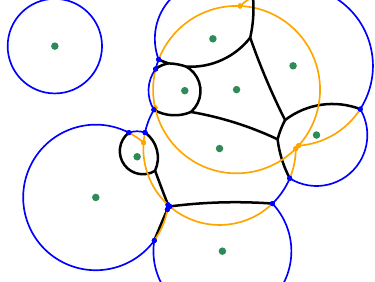}
\caption{A snapshot of the arc expansion for the input shown in
  \Cref{fig:mwvd}. Active arcs that are currently not part of the wavefront
  are drawn in orange.}
\label{fig:act_arc_arr}
\end{figure}

We now describe an event-handling scheme that allows us to trace out
$\VorDiag{S}$ by simulating the expansion of the arcs of $\ArcArr{S,t}$ as $t$
increases, see \Cref{fig:act_arc_arr}. We refer to this process as
\emph{arc expansion}. 

For each site we maintain a search data structure to keep track of all
active arcs during the arc expansion. This \emph{active offset} $\ActOff{i}$
of $s_i$ holds the set of all arcs of $\Off{i}{t}$ which are active at time
$t$ sorted in counter-clockwise angular order around $s_i$, and supports the
following basic operations in time logarithmic in the number of arcs stored:

\begin{itemize}
\item It supports the insertion and deletion of active arcs as well as the
  lookup of their corresponding vertices.
  \item It supports point-location queries, allowing us to identify 
  that active arc within $\ActOff{i}$ which contains a query point 
  $p$ on $\Off{i}{t}$.
\end{itemize}

Every active offset contains at most $2(n-1)$ vertices and, thus, $O(n)$
active arcs. Hence, each such operation on an active offset takes $\BigO(\log
n)$ time in the worst case.

\begin{figure}[htb]
\begin{subfigure}{0.45\textwidth}
  \centering
  \includegraphics[page=13]{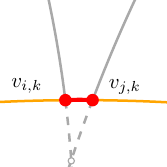}
  \hspace{.1cm}
  \includegraphics[page=14]{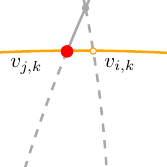}
  \caption{}
  \label{fig:collapse}
\end{subfigure}
\begin{subfigure}{0.45\textwidth}
  \centering
  \includegraphics[page=15]{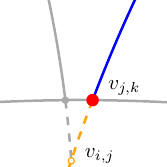}
  \hspace{.1cm}
  \includegraphics[page=16]{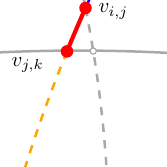}
  \caption{}
  \label{fig:expand}
\end{subfigure}
\caption{(a) A collapse-operation from $\Vertex{i,k}$ to $\Vertex{j,k}$ takes place within $\ActOff{k}$. (b) An expand-operation happens within $\ActOff{j}$ from $\Vertex{j,k}$ to $\Vertex{i,j}$.}
\end{figure}

\begin{figure}[htb]
\begin{subfigure}{0.45\textwidth}
  \centering
  \includegraphics[page=17]{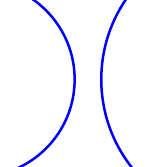}
  \hspace{.1cm}
  \includegraphics[page=18]{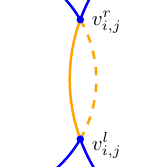}
  \caption{}
  \label{fig:split}
\end{subfigure}
\begin{subfigure}{0.45\textwidth}
  \centering
  \includegraphics[page=19]{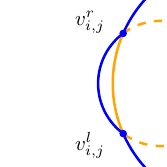}
  \hspace{.1cm}
  \includegraphics[page=20]{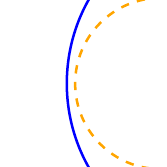}
  \caption{} 
  \label{fig:merge}
\end{subfigure}
  \caption{(a) A split-operation happens when at the time of a collision event. (b) A merge-operation happens at the time of a domination event.}
\end{figure}

Checking and handling the configurations shown in
\Crefrange{fig:collision_arc_ev}{fig:arc_ev} can be done by using only basic
operations within the respective active offsets.  The events themselves are
stored in a priority queue $\Prio$ ordered by the time of their occurrence. If
two events take place simultaneously at the same point then collision events
are prioritized higher than arc events, and arc events have to be handled
before domination events. 
Four auxiliary operations are utilized that allow a more compact description
of this process. Each one takes $\BigO(\log n)$ time.
\begin{itemize}
\item The \emph{collapse-operation} takes place from $\Vertex{i,x}$ to
  $\Vertex{j,k}$ within an active offset $\ActOff{x}$, with $x\in\{j,k\}$, in
  which $\Vertex{i,x}$ and $\Vertex{j,k}$ bound an active arc $\ActArc{x}$
  that is already part of $\ActOff{x}$; see \Cref{fig:collapse}. It determines
  the neighboring active arc $\ActArc{x}'$ of $\ActArc{x}$ that is bounded (on
  one side) by $\Vertex{i,x}$, deletes $\ActArc{x}$ from $\ActOff{x}$, and
  replaces $\Vertex{i,x}$ by $\Vertex{j,k}$ in $\ActArc{x}'$.
\item The counterpart of the collapse-operation is the
  \emph{expand-operation}; see \Cref{fig:expand}. It happens from
  $\Vertex{j,k}$ to $\Vertex{i,x}$ in which $\Vertex{j,k}$ bounds an active
  arc $\ActArc{x}'$ within $\ActOff{x}$. The expansion will either move along
  a currently inactive or an already active portion of the offset circle of
  $s_x$. In the latter case, $\Vertex{j,k}$ is replaced by $\Vertex{i,x}$ in
  $\ActArc{x}'$. In any case, we insert the respective active arc that is
  bounded by $\Vertex{i,x}$ and $\Vertex{j,k}$ into $\ActOff{x}$.
\item A \emph{split-operation} involves two active offsets $\ActOff{i}$ and
  $\ActOff{j}$ as well as a point $p_e$ which is situated within the active
  arcs $\ActArc{i}:=(\Vertex{i,s},\Vertex{i,e})$ and
  $\ActArc{j}:=(\Vertex{j,s'},\Vertex{j,e'})$ within $\ActOff{i}$ and
  $\ActOff{j}$, respectively; see \Cref{fig:split}. Two married vertices
  $\Vertex{i,j}^l$ and $\Vertex{i,j}^r$ are created. Afterwards $\ActArc{i}$
  and $\ActArc{j}$ are removed from $\ActOff{i}$ and $\ActOff{j}$,
  respectively. Two new active arcs $(\Vertex{i,s},\Vertex{i,j}^l)$ and
  $(\Vertex{i,j}^r,\Vertex{i,e})$ are created and inserted into $\ActOff{i}$.
  Furthermore, the three active arcs $(\Vertex{j,e},\Vertex{i,j}^r)$,
  $(\Vertex{i,j}^r, \Vertex{i,j}^l)$, and $(\Vertex{i,j}^l,\Vertex{j,e})$ are
  inserted into $\ActOff{j}$. If $\ActArc{i}$ and $\ActArc{j}$ were
  wavefront arcs then the newly created married vertices coincide with
  wavefront vertices and the newly inserted active arcs except
  $(\Vertex{i,j}^r, \Vertex{i,j}^l)$ are marked as wavefront arcs.
\item During a \emph{merge-operation}, exactly two offset circles interact;
  see \Cref{fig:merge}. The active arcs $\ActArc{i}$ and $\ActArc{j}$ bounded
  by the two corresponding married vertices $\Vertex{i,j}^r$ and
  $\Vertex{i,j}^l$ are removed from $\ActOff{i}$ and $\ActOff{j}$,
  respectively. Additionally, the active arcs $(\Vertex{j,s}, \Vertex{i,j}^r)$
  and $(\Vertex{i,j}^l, \Vertex{j,e})$ that were adjacent to $\ActArc{j}$
  within $\ActOff{j}$ are removed. Finally, a new active arc
  $\ActArc{j}':=(\Vertex{j,s}, \Vertex{j,e})$ is inserted into $\ActOff{j}$.
  If $\ActArc{j}$ was a wavefront arc then $\ActArc{j}'$ is also marked as a
  wavefront arc.
\end{itemize}

Domination events and arc events are easy to detect. The point and time of a
collision is trivial to compute for any pair of offset circles, too.
Unfortunately there is no obvious way to identify those pairs of circles for which this intersection will happen within portions of these offset circles
which will still be active at the time of the collision. Hence, for the rest
of this section we assume that all collisions among all pairs of offset
circles are computed prior to the actual arc expansion. \Cref{lem:events_corr} verifies that
our algorithm correctly simulates the arc expansion.

\begin{lemma}
  For time $t>0$, the arc arrangement $\ArcArr{S,t}$ can be obtained from
  $\ArcArr{S,0}$ by modifying it according to all collision events, domination
  events and arc events that occur till time $t$, in the order in which they
  appear.
\label{lem:events_corr}
\end{lemma}

If the maximum weight of all sites is associated with only one site then there
will be a time $t$ when the offset circle of this site dominates all other
offset circles, i.e., when $\Wavefr{S,t}$ contains only this offset circle as
one active arc. Obviously, at this time no further event can occur and the arc
expansion stops. If multiple sites have the same maximum weight then $\Prio$
can only be empty once $\Wavefr{S,t}$ contains only one loop of active arcs
which all lie on offset circles of these sites and if all wavefront vertices
move along rays to infinity.

\begin{lemma}
  An active arc or active vertex within an active offset
  is identified and marked as a wavefront arc (wavefront vertex, resp.)
  at time $t\ge 0$ if and only if it lies on $\Wavefr{S,t}$.
  \label{lem:active_wavefront}
\end{lemma}

If we allow points in $\R^2$ to
have the same weighted distance to more than three sites then we need to
modify our strategy. In particular, we need to take care of constellations in
which more than three arc events happen simultaneously at the same point. In
such a case it is necessary to carefully choose the sequence in which the
corresponding arc events are handled. More precisely, an arc event may only be
handled (without corrupting the state of the active offsets) whenever the
respective active vertices are considered neighboring within the active
offsets. If the active vertices that participate in an arc event are not
currently neighboring then we can always find an arc event whose active
vertices are neighboring that happens simultaneously at the same location by
walking along the corresponding active offsets. By dealing with the arc events
in this specific order, we generate multiple coinciding Voronoi nodes of
degree three.
Domination events that occur simultaneously at the same point $p_e$ are
processed in increasing order of the weights.
Note that this order can already be
established at the time when an event is inserted into $\Prio$, at no
additional computational cost. Simultaneous multiple collision events at the
same point $p_e$ either involve arcs that are not active or coincide with arc
events. These arc events automatically establish a sorted order of the active
arcs around $p_e$, thus allowing us to avoid an explicit (and time-consuming)
sorting.

\begin{lemma}
  During the arc expansion $\BigO(n^2)$ collision and domination
  events are computed.\label{lem:collisions}
\end{lemma}

We know that collision events create and domination events remove active
vertices (and make them inactive for good).  A collapse of an entire
active-arc triangle causes two vertices to become inactive.  During every
other arc event at least one active vertex becomes inactive, but at the same
time one inactive vertex may become active again.  In order to bound the
number of arc events it is essential to determine how many vertices can be
active and how often a vertex can undergo a \emph{reactivation}, i.e., change
its status from inactive to active. (Note that \Cref{lem:act_point} is not
applicable to a moving vertex since its polar angle does not stay constant.)
We now argue that the total number of reactivations of inactive vertices is
bounded by the number of different vertices that ever were active during
the arc expansion.

\begin{lemma}
  Every reactivation of a moving vertex during an arc event forces another
  moving vertex to become inactive and remain inactive for the rest of the arc
  expansion.
\label{lem:react}
\end{lemma}

\begin{lemma}
  Let $h$ be the number of different vertices that ever were active during the
  arc expansion. Then $\BigO(h)$ arc events can take
  place during the arc expansion. \label{lem:num_arc_events}
\end{lemma}

\begin{theorem}
  The multiplicatively weighted Voronoi diagram $\VorDiag{S}$ of a set $S$ of
  $n$ weighted point sites can be computed in $\BigO(n^2\log n)$ time  and
  $\BigO(n^2)$ space.
\end{theorem}

Additionally, in the appendix we argue that the one-dimensional \ac{mwvd} can be computed efficiently using a wavefront-based strategy.

\begin{theorem}
  \label[theorem]{lem:complexity-1d}
  The multiplicatively weighted Voronoi diagram $\VorDiag{S}$ of a set $S$ of
  $n$ weighted point sites in one dimension can be computed in $\BigO(n \log n)$ time and
  $\BigO(n)$ space.
\end{theorem}

\section{Reducing the Number of Collisions Computed}

Experiments quickly indicate that the vast majority of pairwise collisions
computed a priori never ends up on pairs of active arcs. Furthermore, the
resulting Voronoi diagrams show a quadratic combinatorial complexity only for
contrived input data. We make use of the following results to
determine all collision events in near-linear expected
time. Throughout this section, we assume that for each site $s_i\in S$ the corresponding weight $w(s_i)$ is independently sampled from some probability distribution.

\begin{figure}[!htb]
\centering
\includegraphics[page=14]{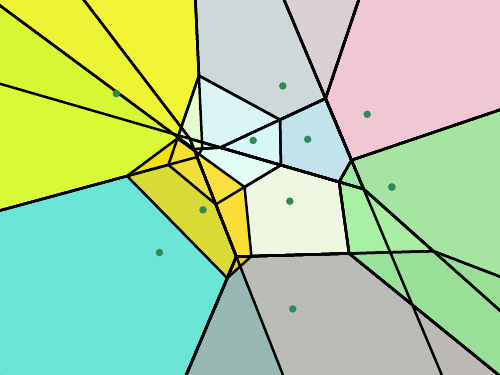}
\caption{The overlay arrangement is generated by inserting the sites ordered by
  decreasing weights.}
\label{fig:overlay}
\end{figure}

\begin{definition}[Candidate Set]
Consider an arbitrary (but fixed) point $q\in\R^2$, and let $s$ be its nearest neighbor in $S$ under the weighted distance. Let $s'\in S\setminus\{s\}$ be another site. Since $s$ is the nearest neighbor of $q$ we know that either $s$ has a higher weight than $s'$ or a smaller Euclidean distance to $q$ than $s'$.  Thus, one can define a \emph{candidate set} for a weighted nearest neighbor of $q$ which consists of all sites $s\in S$ such that all other sites in $S$ either have a smaller weight or a larger Euclidean distance to $q$.
\end{definition}

\begin{lemma}[Har-Peled and Raichel \cite{har2015complexity}]
  For all points $q\in\R^2$, the candidate set for $q$ among $S$ is of size
  $\BigO(\log n)$ with high probability.
\label{lem:cand_set}
\end{lemma}
\begin{lemma}[Har-Peled and Raichel \cite{har2015complexity}]
  Let $K_i$ denote the Voronoi cell of $s_i$ in the unweighted Voronoi diagram
  of the $i$-th suffix $S_i := \{s_i,\ldots,s_n\}$. Let $\OvArr$ denote the
  arrangement formed by the overlay of the regions $K_1,\ldots,K_n$. Then, for
  every face $f$ of $\OvArr$, 
  the candidate set is the same for all points in $f$.
\label{lem:overlay}
\end{lemma}

\Cref{fig:overlay} shows a sample overlay arrangement.  Kaplan et
al.~\cite{kaplan2011overlay} prove that this overlay arrangement has an
expected complexity of $\BigO(n\log n)$. Note that their result is applicable
since inserting the points in sorted order of their randomly chosen weights
corresponds to a randomized insertion. These results allow us to derive better
complexity bounds.

\begin{theorem}[Kaplan et al. \cite{kaplan2011overlay}]
  The expected combinatorial complexity of the overlay of the minimization
  diagrams that arises during a randomized incremental construction of the
  lower envelope of $n$ hyperplanes in $\R^d$, for $d \geq 2$, is
  $\BigO(n^{\lfloor d/2 \rfloor})$, for $d$ even, and $O(n^{\lfloor d/2
    \rfloor}\log n)$, for $d$ odd. The bounds for $d$ even and for $d = 3$ are
  tight in the worst case.
\label{thm:kaplan-overlay}
\end{theorem}

\begin{lemma}
  If a collision event occurs between the offset circles of two sites
  $s_i,s_j\in S$ then there exists at least one candidate set which includes
  both $s_i$ and $s_j$.
\label{lem:valid_coll}
\end{lemma}

\begin{theorem}
  All collision events can be determined in $\BigO(n\log^3 n)$ expected time by computing the overlay arrangement $\OvArr$ of a set $S$ of $n$ input sites.
  \label{thm:overlay}
\end{theorem}

Thus, the number $h$ of vertices created during the arc expansion can be
expected to be bounded by $\BigO(n\log^3 n)$. \Cref{lem:num_arc_events} tells
us that the number of arc events is in $\BigO(h)$. Therefore, $\BigO(n\log^3
n)$ events happen in total.
\begin{theorem}
  A wavefront-based approach allows to compute the multiplicatively weighted
  Voronoi diagram $\VorDiag{S}$ of a set $S$ of $n$ (randomly) weighted point
  sites in expected $\BigO(n\log^4 n)$ time and expected $\BigO(n\log^3 n)$
  space.
\end{theorem}

\section{Extensions} \label{sec:line_segments}

\begin{figure}[htb]
\centering
\includegraphics[page=1]{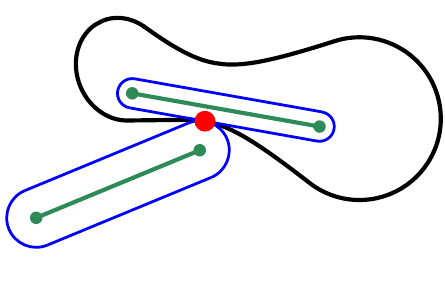}
\hspace{1cm}
\includegraphics[page=2]{bisector}
\caption{An example of a non-piercing (left) as well as a piercing collision event (right).}
\label{fig:sls_bisec}
\end{figure}

Consider a set $S'$ of $n$ disjoint weighted
straight-line segments in $\R^2$. A wavefront propagation among weighted line segments requires us to refine our notion of ``collision''. We call an intersection of two offset circles a \emph{non-piercing collision event} if it marks the initial contact of the two offset circles. That is, it occurs when the first pair of moving vertices appear. We call an intersection of two offset circles a \emph{piercing collision event} if it takes place when two already intersecting offset circles intersect in a third point for the first time; see \Cref{fig:sls_bisec}. In this case, a second pair of moving vertices appear.

\begin{figure}[htb]
\centering
\includegraphics[page=2]{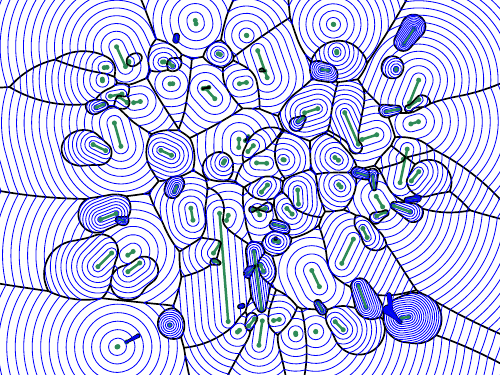}
\caption{The \ac{mwvd} of a set of weighted points and weighted straight-line segments together with a family of wavefronts for equally-spaced points in time.}
\label{fig:sls}
\end{figure}

Hence, a minor modification of our event-based construction scheme is
sufficient to extend it to weighted straight-line segments; see \Cref{fig:sls}. We only need to
check whether a piercing collision event that happens at a point $p_e$ at time
$t_e$ currently is part of $\Wavefr{S',t_e}$. In such a case the two new
vertices as well as the corresponding active arc between them need to be
flagged as part of $\Wavefr{S',t_e}$. 

An extension to additive weights can be integrated easily into our scheme by simply giving every offset circle a head-start of $w_a(s_i)$ at time $t = 0$, where $w_a(s_i)\geq 0$ denotes the real-valued additive weight that is associated with $s_i$.

\section{Experimental Evaluation}

We implemented our full algorithm for multiplicatively weighted points as
input sites\footnote{We do also have a prototype implementation that handles
  both weighted points and weighted straight-line segments. It was used to
  generate \Cref{fig:sls}.}, based on \ac{cgal} and exact
arithmetic\footnote{We have not spent enough time on fine-tuning an
  implementation based on conventional floating-point arithmetic. The obvious
  crux is that inaccurately determined event times (and locations) may corrupt
  the state of the arc arrangement and, thus, cause a variety of errors during
  the subsequent arc expansion.}. In particular, we use \ac{cgal}'s
$\texttt{Arrangement\_2}$ package for computing the overlay arrangement and
its \texttt{Voronoi\_diagram\_2} package for computing unweighted Voronoi
diagrams. The computation of the \ac{mwvd} itself utilizes \ac{cgal}'s
\texttt{Exact\_circular\_kernel\_2} package which is based on the
\texttt{Gmpq} number type. The obvious advantage of using exact number types
is that events are guaranteed to be processed in the right order even if they
occur nearly simultaneously at nearly the same place.  One of the main
drawbacks of exact number types is their memory consumption which is
significantly (and sometimes 
unpredictably) higher than when standard floating-point numbers are used.

We used our implementation for an experimental evaluation and ran our code on
over \num{8000} inputs ranging from \num{256} vertices to \num{500000} vertices. For
all inputs all weights were chosen uniformly at random from the interval $[0,1]$. All tests were carried out with \ac{cgal}~5.0 on
an Intel Core i9-7900X processor clocked at \SI{3.3}{\giga\hertz}.

\begin{figure}[!htbp]
\begin{subfigure}{\textwidth}
\centering
\includegraphics{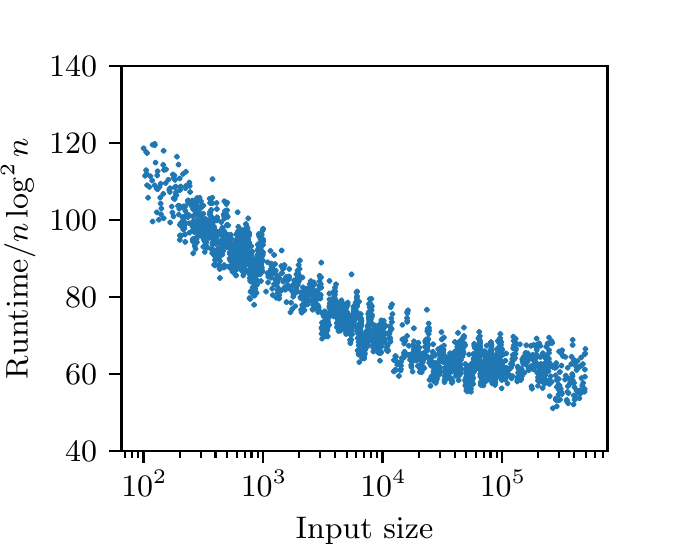}
\includegraphics{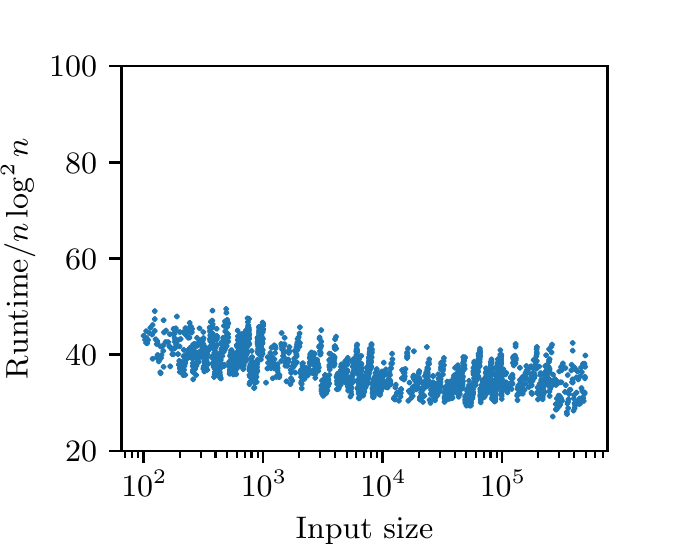}
\caption{
  Left: The overall runtime results for inputs with randomly generated weights and point coordinates. Right: The runtime consumed by the computation of the corresponding overlay arrangements. All runtimes were divided by $n \log^2 n$.}
\label{fig:exp_runtime-random}
\end{subfigure}
\vspace*{1mm}
\begin{subfigure}{\textwidth}
\centering
\includegraphics{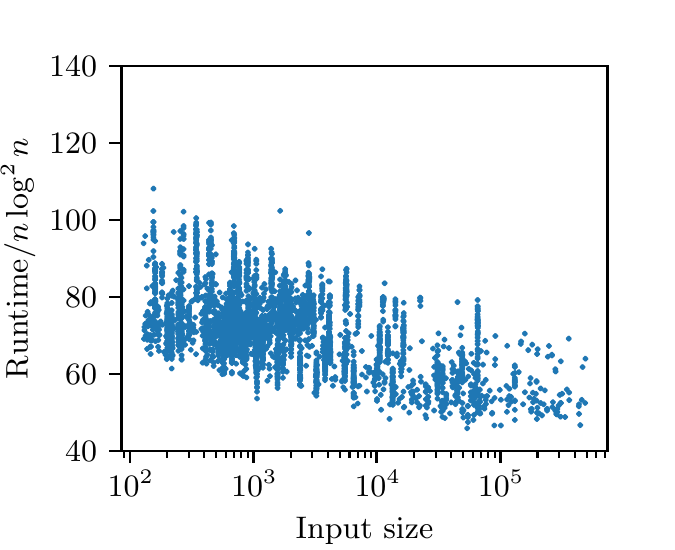}
\caption{
  The overall runtime results for inputs with randomly
  generated weights and vertices of real-world polygons and polygons of the
  \SalzburgDB \cite{Ede*20,SDB20} taken as input points. The runtimes were divided by $n \log^2 n$.}
\label{fig:exp_runtime-real-world}
\end{subfigure}
\vspace*{1mm}
\begin{subfigure}{\textwidth}
\centering
\includegraphics{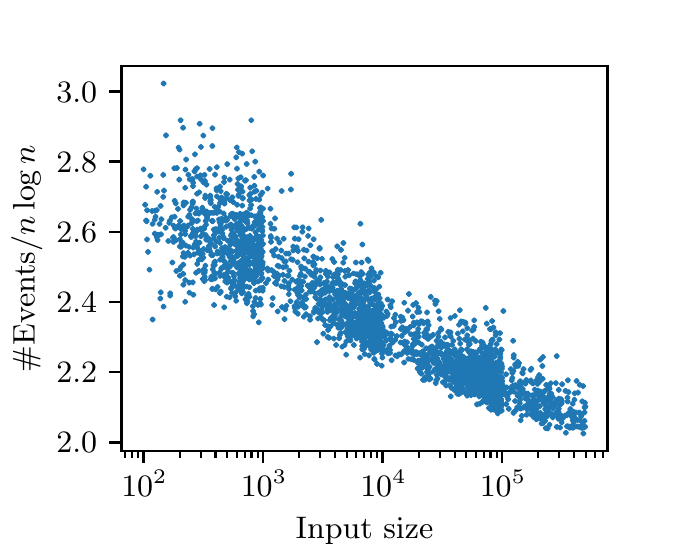}
\includegraphics{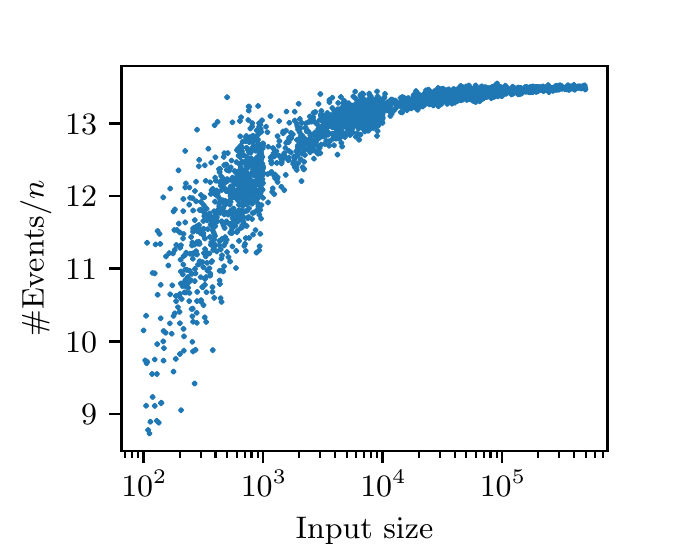}
\caption{The left plot shows the total number of (valid and invalid) collision events (divided by $n \log n$); the right plot shows the number of arc events (divided by $n$) processed during the arc expansion. All point coordinates and weights were generated randomly.} 
\label{fig:exp_events}
\end{subfigure}

\caption{Experimental evaluation.}
\end{figure}

In any case, the number of events is smaller than predicted by the theoretical
analysis.  This is also reflected by our runtime statistics:  In
\Cref{fig:exp_runtime-random,fig:exp_runtime-real-world} the runtime that was consumed by the computation of a
\ac{mwvd} is plotted. 
We ran our tests on two different input classes: The point locations were 
either generated randomly, i.e., they were chosen according to either a uniform or a normal distribution, 
or obtained by taking the vertices of real-world polygons or polygons of the
brand-new \SalzburgDB \cite{Ede*20,SDB20}.
Summarizing, our tests suggest an overall runtime of $\BigO(n \log^2 n)$ for
both input classes. In particular, the actual geometric distribution of the
sites does not have a significant impact on the runtime if the weights are
chosen randomly: For real-world, irregularly distributed sites the runtimes
are scattered more wildly than in the case of uniformly distributed sites, but
they do not increase. 
The numbers of collision events and arc events 
that occurred during the arc expansion are plotted in
\Cref{fig:exp_events}. Our tests suggest that we can expect to see
at most $3 n \log n$ collision events and at most
most $14n$ arc events to occur. Note that the number of arc events forms an
upper bound on the number of Voronoi nodes of the final \ac{mwvd}. That is,
random weights seem to result in a linear combinatorial complexity of the
\ac{mwvd}. 

\begin{figure}[htb]
  \centering
  \includegraphics{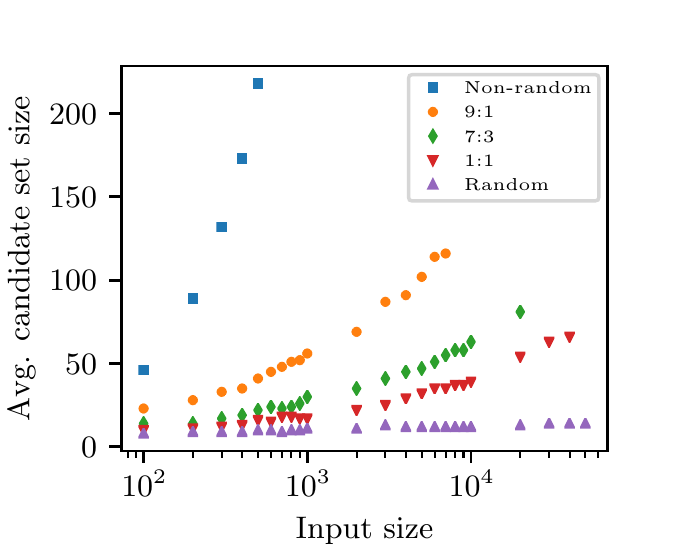}
  \includegraphics{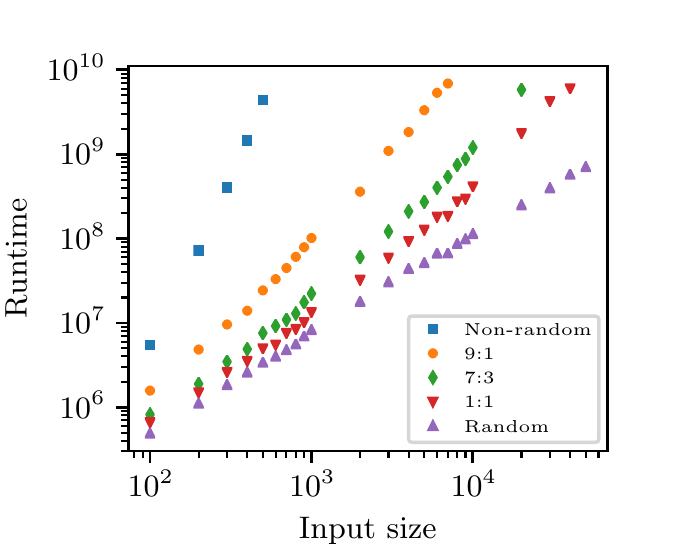}
\caption{The plots show how the average number of candidates (left) and the
  total runtime (right) depend on the weights assigned to the sites.  Each
  marker on the $x$-axes indicates the number $n$ of input sites 
  uniformly distributed within a square.}
  \label{fig:results_non_random}
\end{figure}

It is natural to ask how much these results depend on the randomness of the
weights. To probe this question we set up a second series of experiments: We
sampled points uniformly within a square with side-length $\sqrt{2}$ and
then tested different weights. Let $d(s)$ be the distance of the site $s\in S$
from the center of the square, and let $r(s)$ be a number uniformly
distributed within the interval $[0,1]$. Of course, $0\le d(s)\le 1$. Then we
assign $\nicefrac{\alpha\cdot d(s) + \beta \cdot r(s)}{(\alpha+\beta)}$ as
weight to $s$, with $\alpha$ and $\beta$ being the same arbitrary but fixed
non-negative numbers for all sites of $S$. \Cref{fig:results_non_random} shows
the results obtained for the same sets of points and the
$(\alpha,\beta)$-pairs $(1,0)$, $(9,1)$, $(7,3)$, $(1,1)$ and $(0,1)$. This
test makes it evident that the bounds on the complexities need not hold if the
weights are not chosen randomly, even for a uniform distribution of the
sites. Rather, this may lead to a linear number of candidates per candidate
set and a quadratic runtime complexity, as shown in
\Cref{fig:results_non_random}. 

\section{Conclusion}

\begin{figure}[!htb]
  \centering
  \includegraphics[page=1]{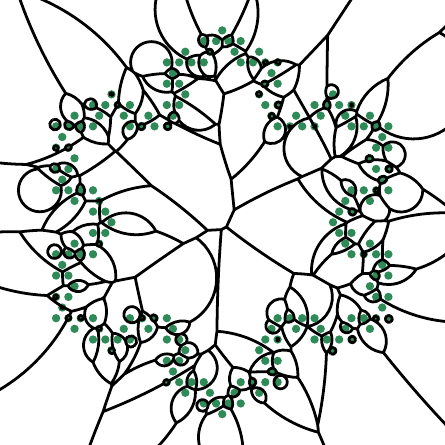}
  \includegraphics[page=1]{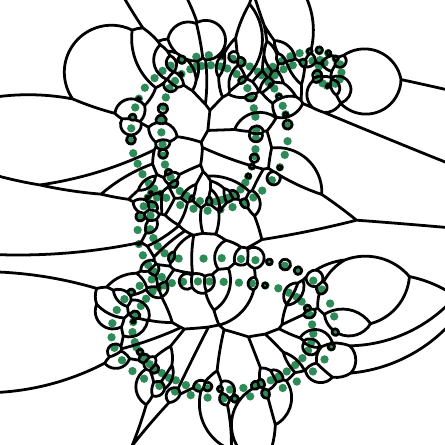}
  \includegraphics[page=1]{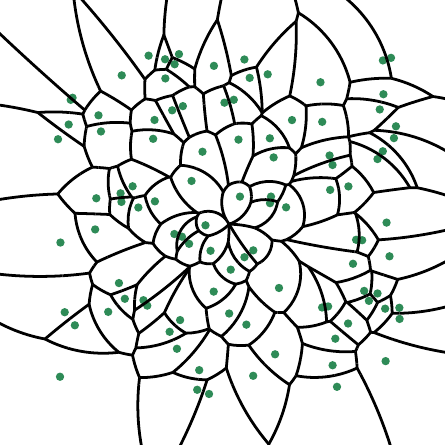}
  \includegraphics[page=2]{koch_snowflake}
  \includegraphics[page=2]{letter_g}
  \includegraphics[page=2]{non-random}
  \caption{Several examples of \acp{mwvd} are shown in the top figures. The bottom figures illustrate a series of uniformly distributed wavefronts that have been derived from the corresponding \acp{mwvd}.}
  \label{fig:mwvd-examples}
\end{figure}

We present a wavefront-like approach for computing the \ac{mwvd} of points and
straight-line segments. Results by Kaplan et al.~\cite{kaplan2011overlay} and
Har-Peled and Raichel \cite{har2015complexity} allow to predict an
$\BigO(n\log^4 n)$ expected time complexity for point sites with random
weights. We also discuss a robust, practical implementation which is based on
\ac{cgal} and exact arithmetic. Extensive tests of our code indicate an
average runtime of $\BigO(n\log^2 n)$ if the sites are weighted randomly. To
the best of our knowledge, there does not exist any other code for computing
\acp{mwvd} that is comparatively fast.  A simple modification of our arc
expansion scheme makes it possible to handle both additive and multiplicative
weights simultaneously. Our code is publicly available on GitHub under
\url{https://github.com/cgalab/wevo}. \Cref{fig:mwvd-examples} shows
several examples of \acp{mwvd} computed by our implementation.

\section*{Acknowledgements}

Work supported by Austrian Science Fund (FWF): Grant P31013-N31.

\bibliography{mwvd}

\newpage

\appendix

\section{Algorithm}

\begin{algorithm}[!htb]
\caption{Compute the \ac{mwvd} $\VorDiag{S}$ of a set of weighted sites $S$.}
\label{alg:basic}
\KwData{A set $S$ of disjoint weighted input sites that are in general position.}
\KwResult{A doubly-connected edge list $\Dcel$ that holds $\VorDiag{S}$.}
Compute all collision events for every pair of sites in $S$ and insert them into the priority queue $\Prio$\;
Insert an active arc that is bounded by two dummy vertices into each active offset that marks the respective offset circle as entirely active\;
Mark all offset circles as being part of the wavefront\;
\While{$\Prio$ is not empty}{
  Pop the next event $e$ from $\Prio$ that occurs at time $t_e>0$ at the point $p_e\in\R^2$\;
  \uIf{$e$ is a collision event}{
    Determine the two offset circles $\Off{i}{t_e}$ and $\Off{j}{t_e}$ that collide\;
    \If{$p_e$ is situated within active arcs $\ActArc{i}$ and $\ActArc{j}$ of $\ActOff{i}$ and $\ActOff{j}$}{
      Split $\ActArc{i}$ and $\ActArc{j}$ at $p_e$\;
      \If{$\ActArc{i}$ and $\ActArc{j}$ are wavefront arcs}{
      Add a new degree-two node $v$ to $\Dcel$ as well as the corresponding half-edges that start at $v$.
      }
    }
  }
  \uElseIf{$e$ is a domination event}{
    Determine $\Off{j}{t_e}$ that dominates $\Off{i}{t_e}$\;
    \If{$p_e$ is situated within active arcs $\ActArc{i}$ and $\ActArc{j}$ of $\ActOff{i}$ and $\ActOff{j}$}{
      Merge the two newly neighboring active arcs within $\ActOff{j}$\;
      \If{$\ActArc{i}$ is a wavefront arc}{
      Add a new degree-two node $v$ to $\Dcel$ and link the corresponding half-edges to $v$.
      }
    }
  }
  \Else{\tcp{$e$ is an arc event.}
    Determine the three vertices $\Vertex{i,j}$, $\Vertex{i,k}$, and $\Vertex{j,k}$ that participate in $e$\;
    \uIf{$\Vertex{i,j}$, $\Vertex{i,k}$, and $\Vertex{j,k}$ are all active}{
      Extract the other two arc events from $\Prio$\ which indicate the disappearances of the remaining two active that form the active-arc triangle\;
      Determine the active arcs $\ActArc{i}$, $\ActArc{j}$, and $\ActArc{k}$ that vanish during $e$. 
      Collapse the active arc $\ActArc{k}$ from $\Vertex{i,k}$ to $\Vertex{j,k}$ within $\ActOff{k}$\;
      Collapse the active arc $\ActArc{j}$ from $\Vertex{i,j}$ to $\Vertex{j,k}$ within $\ActOff{j}$\;
      Remove $\ActArc{i}$ from $\ActOff{i}$\;
    }
    \Else{\tcp{Exactly two vertices are active.}
      Extract the two active vertices $\Vertex{i,x}$ and $\Vertex{j,k}$, with $x\in\{j,k\}$, that bounds the active arc $\ActArc{x}$ whose disappearance is indicated by $e$\;
      Set $y\in\{j,k\}\setminus\{x\}$\;
      Expand the active arc $\ActArc{y}$ from $\Vertex{j,k}$ to $\Vertex{i,y}$ within $\ActOff{y}$\;
      Collapse the active arc $\ActArc{x}$ from $\Vertex{i,x}$ to $\Vertex{j,k}$ within $\ActOff{x}$\; 
      Replace $\Vertex{i,x}$ by $\Vertex{i,y}$ within $\ActOff{i}$\; 
    }
    \If{$\Vertex{i,j}$, $\Vertex{i,k}$, or $\Vertex{j,k}$ are part of $\Wavefr{S,t_e}$}{ 
      Invert the wavefront statuses of $\Vertex{i,j}$, $\Vertex{i,k}$, and $\Vertex{j,k}$\;
      Add a new node $v$ record to $\Dcel$. The respective half-edges are linked to $v$ for every vertex that stops to be a wavefront vertex. Additionally, two new half-edges are produced for each vertex which starts to vertex out a Voronoi edge\;
    }
  }
  Remove the corresponding arc or domination event from $\Prio$ of every active arc that has been modified due to the replacement of one of the vertices that bound it\;
  For each pair of newly neighboring active vertices check if they coincide at a future point in time and insert the respective arc or domination events into $\Prio$\;
}
\end{algorithm}

\newpage

\section{Proofs}

\begin{proof}[Proof of \Cref{lem:act_point}]
  Let $j>i$ be the index of a point site that establishes the inactivity of
  $p_i(\alpha,t)$, and let $t'>t$. The triangle inequality yields
\[
\begin{aligned}
  d(p_i(\alpha,t'),s_j)&\le 
    d(p_i(\alpha,t),s_j)+d(p_i(\alpha,t),p_i(\alpha,t'))
   < w(s_j)\cdot t + (t'-t)\cdot w(s_i) \\
   &\le w(s_j)\cdot t + (t'-t)\cdot w(s_j)=t'\cdot w(s_j), 
\end{aligned}
\]
  which implies that $p_i(\alpha,t')$ lies inside of $\Off{j}{t'}$.
\end{proof}

\begin{proof}[Proof of \Cref{lem:movisects}]
  At least two of the three vertices of an arc event have to be active, and
  the highest-weighted vertex $\Vertex{j,k}$ stays active in any case.  A
  straightforward enumeration of all cases shows that the configurations
  depicted in \cref{fig:arc_ev} are the only configurations possible.
\end{proof}

\begin{proof}[Proof of \Cref{lem:events_corr}]
  All collision events are present in $\Prio$ at the time at which they occur
  as all possible collision events are computed a priori and inserted into
  $\Prio$. Furthermore, whenever two active vertices become neighboring within
  an active offset after a collision, domination, or arc event, we check
  whether they subsequently coincide. Therefore, all domination and arc events
  are also properly detected by our algorithm.

  All offset circles are marked as active at the start time $t=0$. Since all
  sites are distinct, $\ArcArr{S,0}$ corresponds to these (degenerate)
  circles. If no event occurs between the time $t'$ and the time $t''$ then
  also no topological change can occur within $\ArcArr{S,t}$ for $t'<t<t''$.
  Thus, the respective active offsets are initially correct until the first
  event occurs. Assume that $\ArcArr{S,t}$ is correct for all $0 \le t < t_e$
  up until some arbitrary but fixed event $e$ is triggered at time $t_e$.
\begin{itemize}
\item If $e$ is a collision event then two edges of $\ArcArr{S,t_e}$ are
  split at this precise moment. A new edge of $\ArcArr{S,t_e}$ appears at
  $p_e$ whose vertices coincide with the married vertices $\Vertex{i,j}^l$ and
  $\Vertex{i,j}^r$ along $\Off{j}{t_e}$. The split-operation correctly ensures
  that after an collision event all active arcs are still interior-disjoint
  and that their union is equals the new \ac{aa}.
\item If $e$ is a domination event then the updates of the active
  offsets mimic the disappearance of the two edges of $\ArcArr{S,t_e}$ along
  the two offset circles $\Off{i}{t_e}$ and $\Off{j}{t_e}$ whose vertices
  coincide with $\Vertex{i,j}^l$ and $\Vertex{i,j}^r$.
\item Otherwise, $e$ is an arc event at which at least one active arc
  disappears as the three vertices $\Vertex{i,j}$, $\Vertex{i,k}$, and
  $\Vertex{j,k}$ meet in a single point. \Cref{lem:movisects} establishes that
  there exist only six feasible configurations of $\Vertex{i,j}$,
  $\Vertex{i,k}$, $\Vertex{j,k}$ and the corresponding sections along the
  respective offset circles that connect them. If an entire active-arc
  triangle collapses then the isolated active arc along $\Off{i}{t_e}$ which
  is bounded by $\Vertex{i,j}$ and $\Vertex{i,k}$ is consumed by $\Off{j}{t_e}$
  and $\Off{k}{t_e}$. Both $\Vertex{i,j}$ and $\Vertex{i,k}$ become inactive
  during this process. If $e$ marks the disappearance of a single active arc
  then $\Vertex{i,j}$ (or $\Vertex{i,k}$) has been inactive up until $t_e$ and
  becomes active in the process. (Note that the highest-weighted vertex
  $\Vertex{j,k}$ stays active in any case.) Hence, a new active arc appears
  which is bounded by $\Vertex{i,j}$ (or $\Vertex{i,k}$) and $\Vertex{j,k}$, and
  the active arc that is bounded by $\Vertex{i,k}$ (or $\Vertex{i,j}$) and
  $\Vertex{j,k}$ disappears.
\end{itemize}
Thus, the active offsets are correct from $e$ up until the next event takes place. This settles the claim.
\end{proof}

\begin{proof}[Proof of \Cref{lem:active_wavefront}]
  The wavefront status of an active arc or vertex can only change whenever it
  participates in an event. Initially, it is correct for all active arcs and
  vertices. Assume that the wavefront status of all active arcs and vertices
  is correct until some event $e$ takes place at time $t_e$. If $e$ is either
  a collision or domination event then it can be verified easily that the
  split-operation (expand-operation, resp.) appropriately updates the
  wavefront flags of the corresponding active arcs and vertices for every
  feasible configuration of $\Wavefr{S,t_e}$. If $e$ is an arc event then the
  wavefront $\Wavefr{S,t_e}$ changes its topology if and only if at least one
  of the three vertices that are involved coincides with a wavefront vertex.
  The vertices mutually trade places along the corresponding offset circles.
  Therefore, wavefront vertices start to move to the interior of some offset
  circle, and vice versa. Thus, it suffices to invert the wavefront flag of
  all three vertices that participate at an arc event.
\end{proof}

\begin{proof}[Proof of \Cref{lem:react}]
  Let $\Vertex{i,j}$, $\Vertex{i,k}$, and $\Vertex{j,k}$, with $i<j<k$, be the
  three vertices that participate in an arc event $e$ at time $t_e$ such that
  one of them becomes active again, see \Cref{fig:arc_ev}.  Since
  $\Vertex{j,k}$ is the intersection between the two higher-weighted sites it
  is guaranteed to be active before and after the event. So, w.l.o.g.\ assume
  that $\Vertex{i,k}$ gets reactivated. Then $\Vertex{i,j}$ is active before
  and inactive after the event. That is, $\Vertex{i,j}$ is removed from
  $\ArcArr{S,t_e}$ and $\Vertex{i,k}$ is added to $\ArcArr{S,t_e}$.  Let
  $S^*:=\{s_i,s_j,s_k\}$. Then this implies that $\Vertex{i,j}$ would also be
  removed from $\ArcArr{S^*,t_e}$ and $\Vertex{i,k}$ added to
  $\ArcArr{S^*,t_e}$, as illustrated in \Cref{fig:arc_ev}.

  Since $\Vertex{i,k}$ is reactivated we know that it had been active in
  $\ArcArr{S,t'}$ at some time $t'<t$. This implies that it had been active
  also in $\ArcArr{S^*,t'}$, and that it had become inactive relative to $S^*$
  during some further arc event $e'$ that took place at time $t' < t_{e'} <
  t_e$. Now recall that every arc event corresponds to a Voronoi node.  Since
  $\VorDiag{S^*}$ has at most two nodes, the time $t_e$ marks the last time at
  which an arc event may take place during the construction of
  $\VorDiag{S^*}$. Hence, $\Vertex{i,j}$ becomes inactive relative to $S^*$ at
  time $t_e$ and is forced to stay inactive. In particular, it cannot belong
  to $\ArcArr{S^*,t}$ for any $t>t_e$.  Thus, it cannot belong to
  $\ArcArr{S,t}$ either but stays inactive for the entire rest of the arc
  expansion.
\end{proof}

\begin{proof}[Proof of \Cref{lem:num_arc_events}]
  At each collision event a new pair of active vertices is generated, while
  domination events remove active vertices for good and do not generate new
  vertices. At every arc event at least one active vertex is deactivated and
  no new vertex is generated.  Hence, the number $h$ is bounded by the number
  of collision events.  \Cref{lem:react} tells us that the reactivation of one
  vertex during an arc event can be charged to another vertex which becomes
  inactive at that time and remains inactive for the entire rest of the arc
  expansion. Thus, the overall number of arc events is in $\BigO(h)$.
\end{proof}

\begin{proof}[Proof of \Cref{lem:valid_coll}]
  The unweighted Voronoi region $\mathcal{VR}(s_i,S_i)$ of $s_i$ relative to
  $S_i$ bounds the maximum extent of the (weighted) Voronoi region
  $\mathcal{VR}_w(s_i,S)$ in the final multiplicatively weighted Voronoi
  diagram $\mathcal{VD}_w(S)$. Thus, for all $t\in\R^+$, all active arcs of
  $\Off{i}{t}$ are restricted to $\mathcal{VR}(s_i,S_i)$.  Therefore, $s_i$ is
  present in all candidate sets in which active arcs of $\Off{i}{t}$ can be 
  situated. Of course, the same argument holds for $s_j$.
\end{proof}

\begin{proof}[Proof of \Cref{thm:overlay}]
  According to Har-Peled and Raichel \cite{har2015complexity}, the computation
  of $\OvArr$ takes $\BigO(n\log^2 n)$ expected time.
  A collision event can only occur between sites that are within the same
  candidate set. Thus, we iterate over all faces of $\OvArr$. Let $C$ be the
  candidate set of such a face. We insert a collision event into the event
  queue $\Prio$ for every pair of sites in $C$. \Cref{lem:cand_set} tells us
  that we can expect to insert $\BigO(\log^2 n)$ collision events for $C$.
  Since $\OvArr$ has an expected number of at most $\BigO(n\log n)$ faces, as
  established in \Cref{thm:kaplan-overlay}, the expected total number of collision
  events that is added to $\Prio$ is bounded by $\BigO(n\log^3 n)$.  Finally,
  \Cref{lem:valid_coll} ensures that this approach does indeed detect all
  collision events.
\end{proof}

\section{The One-Dimensional Weighted Voronoi Diagram}
\label{sec:one-dimensional}

Consider a set $S$ of $n$ point sites in $\R$ where every site $s_i$ has a
strictly positive weight $w(s_i)\in \R^+$.  Aurenhammer \cite{aurenhammer1986one}
shows that $\VorDiag{S}$ has a linear combinatorial complexity by modeling it
as the lower envelope of wedges in $\R^2$. The actual Voronoi diagram is
computed in $\BigO(n \log n)$ time by means of divide\&conquer and a plane
sweep.

We now apply our wavefront propagation to establish the linear combinatorial
complexity of $\VorDiag{S}$ and to compute it in $\BigO(n \log n)$ time. In
this one-dimensional setting, an offset circle of $s_i$ consists of two moving
\emph{offset points} $p^l_i$ and $p^r_i$, where the offset point $p^l_i$ lies
left of $s_i$ and moves leftwards, and $p^r_i$ lies right of $s_i$ and moves
rightwards.
Similar to \Cref{sec:event_based}, the following two events can occur during
the wavefront propagation: 
\begin{itemize}
  \item A \emph{collision} event occurs if two offset points that
    move in opposite directions coincide.
  \item A \emph{domination event} occurs if two offset points that
    move in the same direction coincide.
\end{itemize}
We get no arc event since there are no offset arcs.

All offset points are kept in sorted left-to-right order in a doubly-linked
list $\Sort$. In particular, every offset point knows its neighbor to the left
and to the right.  Furthermore, every offset point holds a flag that indicates
whether it currently lies on the wavefront. Initially, we compute all
collision events between neighboring offset points, and insert them into
$\Prio$.  Furthermore, all offset points are flagged to lie on the wavefront.

We now explain the handling of an event. Let $s_i$ and $s_j$ be the sites
whose offset points are involved in this event. W.l.o.g., we assume that $s_i$
is left of $s_j$. If $w(s_i)<w(s_j)$ then the following two events are possible.
\begin{itemize}
\item Collision event: If both $p_i^r$ and $p_j^l$ are on the wavefront then
  we have discovered a new Voronoi node. 
\item Domination event: If $p_i^l$ lies on the wavefront then a new Voronoi
  node has been discovered and $p_j^l$ now is flagged to lie on the wavefront.
\end{itemize}
For both events, the offset point of $s_i$ is deleted from $\Sort$ and the collision/domination
event that is currently associated with it is deleted from $\Prio$.
Furthermore, we update the left neighbor of $p_j^l$, and insert the respective
collision or domination event into $\Prio$.

If $w(s_i)=w(s_j)$ then a domination event cannot occur. In the case of a
collision event both $s_i$ and $s_j$ are deleted from $\Sort$, and the events
currently associated with them are deleted from $\Prio$.

\begin{proof}[Proof of \Cref{lem:complexity-1d}]
  The sites send out $2n$ moving offset points which are inserted in
  left-to-right order in $\Sort$. Furthermore we get $n-1$ initial
  collision events that are inserted into $\Prio$. 
  At every event at least one moving offset point is deleted (and cannot
  trigger subsequent events), and at most one collision/domination event is replaced
  by another collision/domination event in $\Prio$. Overall exactly $2n -2$
  moving offset points can be and will be deleted. Hence we get a linear
  number of events and a constant number of updates of $\Prio$ and $\Sort$ per
  event. This guarantees a total runtime of $\BigO(n \log n)$ time.
\end{proof}

\end{document}